\def \inparg{\leftskip = 40pt\rightskip = 40pt}
\def \outparg{\leftskip = 0 pt\rightskip = 0pt}

\def\npb{{Nucl.\ Phys.\ }{\bf B}}
\def\plb{{Phys.\ Lett.\ }{ \bf B}}

\def\prd{{Phys.\ Rev.\ }{\bf D}}

\def\prl{Phys.\ Rev.\ Lett.\ }

\def\ov #1{\overline{#1}}
\def\wt #1{\widetilde{#1}}
\def\Tr{\mathop{\rm Tr}}

\def\th{{\theta}}
\def\bth{{\overline{\theta}}}

\def\frak#1#2{{\textstyle{{#1}\over{#2}}}}
\def\nhalf{${\cal N} = \frak{1}{2}$}
\def\none{${\cal N} = 1$}

\def\Ahat{\hat A}
\def\Chat{\hat C}

\def\frak#1#2{{\textstyle{{#1}\over{#2}}}}

\def\go{\rightarrow}
\def\lambdabar{\bar\lambda}
\def\lambdahatbar{\bar{\hat\lambda}}
\def\lambdahat{\hat\lambda}

\def\Dhat{\hat D}
\def\Ftil{\tilde F}
\def\Fhat{\hat F}

\def\sigmabar{\bar\sigma}
\def\phibar{\bar\phi}
\def\phitilbar{\bar{\tilde{\phi}}}
\def\psitilbar{\bar{\tilde{\psi}}}

\def\psibar{\bar\psi}
\def\Fbar{\bar F}

\def\phitil{\tilde\phi}
\def\psitil{\tilde\psi}
\def\Ncal{{\cal N}}
\def\Ftil{\tilde F}
\def\alphadot{\dot\alpha}
\def\betadot{\dot\beta}

\def\pa{\partial}

\input harvmac
\input epsf
%
\newbox\hdbox%
\newcount\hdrows%
\newcount\multispancount%
\newcount\ncase%
\newcount\ncols
\newcount\nrows%
\newcount\nspan%
\newcount\ntemp%
\newdimen\hdsize%
\newdimen\newhdsize%
\newdimen\parasize%
\newdimen\spreadwidth%
\newdimen\thicksize%
\newdimen\thinsize%
\newdimen\tablewidth%
\newif\ifcentertables%
\newif\ifendsize%
\newif\iffirstrow%
\newif\iftableinfo%
\newtoks\dbt%
\newtoks\hdtks%
\newtoks\savetks%
\newtoks\tableLETtokens%
\newtoks\tabletokens%
\newtoks\widthspec%
%
%
%
%
\tableinfotrue%
\catcode`\@=11
%
%
\def\tstrut{\vrule height3.1ex depth1.2ex width0pt}%
\def\and{\char`\&}
\def\tablerule{\noalign{\hrule height\thinsize depth0pt}}%
\thicksize=1.5pt
\thinsize=0.6pt
\def\thickrule{\noalign{\hrule height\thicksize depth0pt}}%
\def\ctr#1{\hfil\ #1\hfil}%
%
%
%
%
\tablewidth=-\maxdimen%
\spreadwidth=-\maxdimen%
\def\tabskipglue{0pt plus 1fil minus 1fil}%
%
%
\centertablestrue%
%
%
%
%
\parasize=4in%
\gdef\ARGS{########}
\gdef\headerARGS{####}
\def\@mpersand{&}
{\catcode`\|=13
\gdef\letbarzero{\let|0}
\gdef\letbartab{\def|{&&}}%
\gdef\letvbbar{\let\vb|}%
}
{\catcode`\&=4
\def\ampskip{&\omit\hfil&}
\catcode`\&=13
\let&0
\xdef\letampskip{\def&{\ampskip}}%
\gdef\letnovbamp{\let\novb&\let\tab&}
}
\def\begintable{
   \begingroup%
   \catcode`\|=13\letbartab\letvbbar%
   \catcode`\&=13\letampskip\letnovbamp%
   \def\multispan##1{
      \omit \mscount##1%
      \multiply\mscount\tw@\advance\mscount\m@ne%
      \loop\ifnum\mscount>\@ne \sp@n\repeat%
   }
   \def\|{%
      &\omit\widevline&%
   }%
   \ruledtable
}
\long\def\ruledtable#1\endtable{%
%
%
%
   \offinterlineskip
   \tabskip 0pt
   \def\widevline{\vrule width\thicksize}
   \def\endrow{\@mpersand\omit\hfil\crnorm\@mpersand}%
   \def\crthick{\@mpersand\crnorm\thickrule\@mpersand}%
   \def\crthickneg##1{\@mpersand\crnorm\thickrule
          \noalign{{\skip0=##1\vskip-\skip0}}\@mpersand}%
   \def\crnorule{\@mpersand\crnorm\@mpersand}%
   \def\crnoruleneg##1{\@mpersand\crnorm
          \noalign{{\skip0=##1\vskip-\skip0}}\@mpersand}%
   \let\nr=\crnorule
   \def\endtable{\@mpersand\crnorm\thickrule}%
   \let\crnorm=\cr
%
%
   \edef\cr{\@mpersand\crnorm\tablerule\@mpersand}%
   \def\crneg##1{\@mpersand\crnorm\tablerule
          \noalign{{\skip0=##1\vskip-\skip0}}\@mpersand}%
   \let\ctneg=\crthickneg
   \let\nrneg=\crnoruleneg
   \the\tableLETtokens
%
%
   \tabletokens={&#1}
%
%
   \countROWS\tabletokens\into\nrows%
   \countCOLS\tabletokens\into\ncols%
%
%
   \advance\ncols by -1%
   \divide\ncols by 2%
   \advance\nrows by 1%
%
%
   \iftableinfo %
      \immediate\write16{[Nrows=\the\nrows, Ncols=\the\ncols]}%
   \fi%
%
%
   \ifcentertables
      \ifhmode \par\fi
      \line{
      \hss
   \else %
      \hbox{%
   \fi
      \vbox{%
         \makePREAMBLE{\the\ncols}
         \edef\next{\preamble}
         \let\preamble=\next
         \makeTABLE{\preamble}{\tabletokens}
      }
      \ifcentertables \hss}\else }\fi
   \endgroup
   \tablewidth=-\maxdimen
   \spreadwidth=-\maxdimen
}
\def\makeTABLE#1#2{
   {
   \let\ifmath0
   \let\header0
   \let\multispan0
%
%
   \ncase=0%
   \ifdim\tablewidth>-\maxdimen \ncase=1\fi%
   \ifdim\spreadwidth>-\maxdimen \ncase=2\fi%
   \relax
%
   \ifcase\ncase %
      \widthspec={}%
   \or %
      \widthspec=\expandafter{\expandafter t\expandafter o%
                 \the\tablewidth}%
   \else %
      \widthspec=\expandafter{\expandafter s\expandafter p\expandafter r%
                 \expandafter e\expandafter a\expandafter d%
                 \the\spreadwidth}%
   \fi %
   \xdef\next{
      \halign\the\widthspec{%
      #1
      \noalign{\hrule height\thicksize depth0pt}
      \the#2\endtable
%
      }
   }
   }
   \next
}
\def\makePREAMBLE#1{
   \ncols=#1
   \begingroup
   \let\ARGS=0
   \edef\xtp{\widevline\ARGS\tabskip\tabskipglue%
   &\ctr{\ARGS}\tstrut}
   \advance\ncols by -1
   \loop
      \ifnum\ncols>0 %
      \advance\ncols by -1%
      \edef\xtp{\xtp&\vrule width\thinsize\ARGS&\ctr{\ARGS}}%
   \repeat
   \xdef\preamble{\xtp&\widevline\ARGS\tabskip0pt%
   \crnorm}
   \endgroup
}
\def\countROWS#1\into#2{
   \let\countREGISTER=#2%
   \countREGISTER=0%
   \expandafter\ROWcount\the#1\endcount%
}%
\def\ROWcount{%
   \afterassignment\subROWcount\let\next= %
}%
\def\subROWcount{%
   \ifx\next\endcount %
      \let\next=\relax%
   \else%
      \ncase=0%
      \ifx\next\cr %
         \global\advance\countREGISTER by 1%
         \ncase=0%
      \fi%
      \ifx\next\endrow %
         \global\advance\countREGISTER by 1%
         \ncase=0%
      \fi%
      \ifx\next\crthick %
         \global\advance\countREGISTER by 1%
         \ncase=0%
      \fi%
      \ifx\next\crnorule %
         \global\advance\countREGISTER by 1%
         \ncase=0%
      \fi%
      \ifx\next\crthickneg %
         \global\advance\countREGISTER by 1%
         \ncase=0%
      \fi%
      \ifx\next\crnoruleneg %
         \global\advance\countREGISTER by 1%
         \ncase=0%
      \fi%
      \ifx\next\crneg %
         \global\advance\countREGISTER by 1%
         \ncase=0%
      \fi%
      \ifx\next\header %
         \ncase=1%
      \fi%
      \relax%
      \ifcase\ncase %
         \let\next\ROWcount%
      \or %
         \let\next\argROWskip%
      \else %
      \fi%
   \fi%
   \next%
}
\def\counthdROWS#1\into#2{%
\dvr{10}%
   \let\countREGISTER=#2%
   \countREGISTER=0%
\dvr{11}%
\dvr{13}%
   \expandafter\hdROWcount\the#1\endcount%
\dvr{12}%
}%
\def\hdROWcount{%
   \afterassignment\subhdROWcount\let\next= %
}%
\def\subhdROWcount{%
   \ifx\next\endcount %
      \let\next=\relax%
   \else%
      \ncase=0%
      \ifx\next\cr %
         \global\advance\countREGISTER by 1%
         \ncase=0%
      \fi%
      \ifx\next\endrow %
         \global\advance\countREGISTER by 1%
         \ncase=0%
      \fi%
      \ifx\next\crthick %
         \global\advance\countREGISTER by 1%
         \ncase=0%
      \fi%
      \ifx\next\crnorule %
         \global\advance\countREGISTER by 1%
         \ncase=0%
      \fi%
      \ifx\next\header %
         \ncase=1%
      \fi%
\relax%
      \ifcase\ncase %
         \let\next\hdROWcount%
      \or%
         \let\next\arghdROWskip%
      \else %
      \fi%
   \fi%
   \next%
}%
{\catcode`\|=13\letbartab
\gdef\countCOLS#1\into#2{%
   \let\countREGISTER=#2%
   \global\countREGISTER=0%
   \global\multispancount=0%
   \global\firstrowtrue
   \expandafter\COLcount\the#1\endcount%
   \global\advance\countREGISTER by 3%
   \global\advance\countREGISTER by -\multispancount
}%
\gdef\COLcount{%
   \afterassignment\subCOLcount\let\next= %
}%
{\catcode`\&=13%
\gdef\subCOLcount{%
   \ifx\next\endcount %
      \let\next=\relax%
   \else%
      \ncase=0%
      \iffirstrow
         \ifx\next& %
            \global\advance\countREGISTER by 2%
            \ncase=0%
         \fi%
         \ifx\next\span %
            \global\advance\countREGISTER by 1%
            \ncase=0%
         \fi%
         \ifx\next| %
            \global\advance\countREGISTER by 2%
            \ncase=0%
         \fi
         \ifx\next\|
            \global\advance\countREGISTER by 2%
            \ncase=0%
         \fi
         \ifx\next\multispan
            \ncase=1%
            \global\advance\multispancount by 1%
         \fi
         \ifx\next\header
            \ncase=2%
         \fi
         \ifx\next\cr       \global\firstrowfalse \fi
         \ifx\next\endrow   \global\firstrowfalse \fi
         \ifx\next\crthick  \global\firstrowfalse \fi
         \ifx\next\crnorule \global\firstrowfalse \fi
         \ifx\next\crnoruleneg \global\firstrowfalse \fi
         \ifx\next\crthickneg  \global\firstrowfalse \fi
         \ifx\next\crneg       \global\firstrowfalse \fi
      \fi
\relax
      \ifcase\ncase %
         \let\next\COLcount%
      \or %
         \let\next\spancount%
      \or %
         \let\next\argCOLskip%
      \else %
      \fi %
   \fi%
   \next%
}%
\gdef\argROWskip#1{%
   \let\next\ROWcount \next%
}
\gdef\arghdROWskip#1{%
   \let\next\ROWcount \next%
}
\gdef\argCOLskip#1{%
   \let\next\COLcount \next%
}
}
}
\def\spancount#1{
   \nspan=#1\multiply\nspan by 2\advance\nspan by -1%
   \global\advance \countREGISTER by \nspan
   \let\next\COLcount \next}%
\def\dvr#1{\relax}%
\def\header#1{%
\dvr{1}{\let\cr=\@mpersand%
\hdtks={#1}%
\counthdROWS\hdtks\into\hdrows%
\advance\hdrows by 1%
\ifnum\hdrows=0 \hdrows=1 \fi%
\dvr{5}\makehdPREAMBLE{\the\hdrows}%
\dvr{6}\getHDdimen{#1}%
{\parindent=0pt\hsize=\hdsize{\let\ifmath0%
\xdef\next{\valign{\headerpreamble #1\crnorm}}}\dvr{7}\next\dvr{8}%
}%
}\dvr{2}}
\def\makehdPREAMBLE#1{
\dvr{3}%
\hdrows=#1
{
\let\headerARGS=0%
\let\cr=\crnorm%
\edef\xtp{\vfil\hfil\hbox{\headerARGS}\hfil\vfil}%
\advance\hdrows by -1
\loop
\ifnum\hdrows>0%
\advance\hdrows by -1%
\edef\xtp{\xtp&\vfil\hfil\hbox{\headerARGS}\hfil\vfil}%
\repeat%
\xdef\headerpreamble{\xtp\crcr}%
}
\dvr{4}}
\def\getHDdimen#1{%
\hdsize=0pt%
\getsize#1\cr\end\cr%
}
\def\getsize#1\cr{%
\endsizefalse\savetks={#1}%
\expandafter\lookend\the\savetks\cr%
\relax \ifendsize \let\next\relax \else%
\setbox\hdbox=\hbox{#1}\newhdsize=1.0\wd\hdbox%
\ifdim\newhdsize>\hdsize \hdsize=\newhdsize \fi%
\let\next\getsize \fi%
\next%
}%
\def\lookend{\afterassignment\sublookend\let\looknext= }%
\def\sublookend{\relax%
\ifx\looknext\cr %
\let\looknext\relax \else %
   \relax
   \ifx\looknext\end \global\endsizetrue \fi%
   \let\looknext=\lookend%
    \fi \looknext%
}%
%
%
\def\tablelet#1{%
   \tableLETtokens=\expandafter{\the\tableLETtokens #1}%
}%
\catcode`\@=12

{\nopagenumbers
\line{\hfil LTH 708}
\line{\hfil hep-th/0607194}
\vskip .5in
\centerline{\titlefont One-loop renormalisation of massive}
\centerline{\titlefont $\Ncal=\frak12$ supersymmetric gauge theory}
\vskip 1in
\centerline{\bf I.~Jack, D.R.T.~Jones and L.A.~Worthy}
\bigskip
\centerline{\it Department of Mathematical Sciences,  
University of Liverpool, Liverpool L69 3BX, U.K.}
\vskip .3in
We construct the general $\Ncal=\frak12$ 
supersymmetric gauge theory coupled to massive chiral matter, and show
that it is renormalisable at one loop.  
\Date{July 2006}}

\newsec{Introduction} $\Ncal=\frak12$ supersymmetric theories (i.e.
theories defined on non-anticommutative superspace) have recently attracted
a good deal of attention.\ref\ferr{S.~Ferrara and 
M.A.~Lledo, JHEP 0005 (2008) 008}\nref\klemm{D.~Klemm, S.~Penati and 
L.~Tamassia, Class. Quant. Grav.
20 (2003) 2905}\nref\seib{N.~Seiberg, JHEP
{\bf 0306} (2003) 010}--\ref\araki{T.~Araki, K.~Ito and  A. Ohtsuka,
\plb573 (2003) 209}. Such theories are non-hermitian and 
only have half the supersymmetry of the corresponding $\Ncal=1$
theory. 
These theories are not power-counting  renormalisable\foot{See
Refs.~\ref\britto{R.~Britto, B.~Feng and S.-J.~Rey, 
JHEP 0307 (2003) 067; JHEP 0308 (2003) 001}\ref\terash{S. Terashima 
and J-T Yee, JHEP {\bf 0312} 
(2003) 053} for other discussions of the ultra-violet properties of these
theories.} but it has been
argued\ref\gris{M.T.~Grisaru, S.~Penati and  A.~Romagnoni, JHEP {\bf
0308} (2003) 003\semi
R.~Britto and B.~Feng, \prl91 (2003) 201601\semi
A.~Romagnoni, JHEP {\bf 0310} (2003) 016}\nref\lunin{O.~Lunin 
and S.-J. Rey, JHEP  {\bf 0309}
(2003) 045}\nref\alish{
M.~Alishahiha, A.~Ghodsi and N.~Sadooghi, \npb691
(2004) 111}--\ref\berrey{D.~Berenstein and S.-J.~Rey, \prd68 (2003) 121701}
 that they are in  fact nevertheless
renormalisable, in the sense that only a finite number of additional
terms need to be added to the lagrangian to absorb divergences to all
orders. In previous work we have confirmed this renormalisability at
the one-loop level, at the same time revealing unexpected subtleties.
We showed that divergent 
gauge non-invariant terms are generated, which however may be removed by a 
divergent field redefinition\ref\jjw{I.~Jack, D.R.T.~Jones
and L.A.~Worthy, \plb611 (2005) 199}; and that in the case of $\Ncal=\frak12$ 
supersymmetry with chiral matter (in the fundamental 
representation)\ref\jjwa{I.~Jack, D.R.T.~Jones
and L.A.~Worthy, \prd72 (2005) 065002 }\ the joint 
requirements of
renormalisability and $\Ncal=\frak12$ supersymmetry impose the choice of
gauge group $SU(N)\otimes U(1)$ (rather than $U(N)$ or $SU(N)$). 
It is interesting to compare our results with those obtained using 
superfields. The
authors of Ref.~\ref\penrom{
S.~Penati and A.~Romagnoni, JHEP {\bf 0502} (2005) 064}
obtained the one loop effective action for pure $\Ncal=\frak12$
supersymmetry using a superfield formalism. Although they found
divergent contributions which broke supergauge invariance, their final
result was gauge-invariant without the need for any redefinition. In
subsequent work\ref\gpr{M.T.~Grisaru, S.~Penati and A.~Romagnoni,
JHEP {\bf 0602} (2006) 043} it was shown that the $\Ncal=\frak12$ 
superfield action requires modification to ensure renormalisability, which
is consistent with our findings in the component formulation\jjwa.
In Ref.~\jjwa\
there was no superpotential for the chiral matter. In the present work we
continue to examine the $\Ncal=\frak12$ supersymmetric 
theory with chiral matter in the fundamental representation and explore
the consequences of adding superpotential terms, which consist 
of mass terms for the chiral and antichiral fields (linking the 
fundamental and antifundamental representations).  

The action for an $\Ncal=\frak12$ supersymmetric $SU(N)\otimes U(1)$ gauge 
theory coupled to 
chiral matter (with no superpotential) is given by\jjwa
\eqn\lagranb{\eqalign{
S=&\int d^4x
\Bigl[-\frak14F^{\mu\nu A}F^A_{\mu\nu}-i\lambdabar^A\sigmabar^{\mu}
(D_{\mu}\lambda)^A+\frak12D^AD^A\cr
&-\frak12iC^{\mu\nu}d^{ABC}e^{ABC}F^A_{\mu\nu}\lambdabar^B\lambdabar^C\cr
&+\frak18g^2|C|^2d^{abe}d^{cde}(\lambdabar^a\lambdabar^b)
(\lambdabar^c\lambdabar^d)
+\frak{1}{4N}\frak{g^4}{g_0^2}|C|^2(\lambdabar^a\lambdabar^a)
(\lambdabar^b\lambdabar^b)\cr
&+\Bigl\{
\Fbar F -i\psibar\sigmabar^{\mu}D_{\mu}\psi-D^{\mu}\phibar D_{\mu}\phi\cr
&+\phibar \Dhat\phi +
i\sqrt2(\phibar \lambdahat\psi-\psibar\lambdahatbar\phi)\cr
&+\sqrt2C^{\mu\nu}D_{\mu}\phibar\lambdahatbar\sigmabar_{\nu}\psi
+iC^{\mu\nu}\phibar \Fhat_{\mu\nu}F
-\frak{1}{4}|C|^2\phibar \lambdahatbar\lambdahatbar
F\cr
&+\frak{1}{N}\gamma_1g_0^2|C|^2(\lambdabar^a\lambdabar^a)
(\lambdabar^0\lambdabar^0)\cr
&-\gamma_2 C^{\mu\nu}g\left(
\sqrt2D_{\mu}\phibar\lambdabar^aR^a\sigmabar_{\nu}\psi+\sqrt2\phibar
\lambdabar^aR^a
\sigmabar_{\nu}D_{\mu}\psi+i\phibar F^a_{\mu\nu}R^aF\right)\cr
&-\gamma_3 C^{\mu\nu}g_0\left(
\sqrt2D_{\mu}\phibar\lambdabar^0R^0\sigmabar_{\nu}\psi+\sqrt2\phibar
\lambdabar^0R^0
\sigmabar_{\nu}D_{\mu}\psi+i\phibar F^0_{\mu\nu}R^0F\right)\cr
&+(\phi\go\phitil,\psi\go\psitil,F\go\Ftil,R^A\go -(R^A)^*,C^{\mu\nu}\go
-C^{\mu\nu})\Bigr\}\Bigr],\cr}} 
where $\gamma_{1-3}$ are constants, and 
\eqn\afdef{\eqalign{
D_{\mu}\phi=&\pa_{\mu}\phi+i\Ahat_{\mu}\phi,\cr
(D_{\mu}\lambda)^A=&\pa_{\mu}\lambda^A-gf^{ABC}A_{\mu}^B\lambda^C,\cr
\quad F_{\mu\nu}^A=&\pa_{\mu}A_{\nu}^A-\pa_{\nu}A_{\mu}^A-gf^{ABC}
A_{\mu}^BA_{\nu}^C,\cr}}
(with similar expressions for $D_{\mu}\phitil$, $D_{\mu}\psi$,
$D_{\mu}\psitil$). Here
\eqn\hatdefs{
\Ahat_{\mu}=\Ahat_{\mu}^AR^A=gA_{\mu}^aR^a+g_0A_{\mu}^0R^0,}
with similar definitions for $\lambdahat$, $\Dhat$, $\Fhat_{\mu\nu}$.
We also have
\eqn\etensor{
e^{abc}=g,\quad e^{a0b}=e^{ab0}=e^{000}=g_0,\quad e^{0ab}={g^2\over{g_0}}.}
We include a multiplet $\{\phi,\psi,F\}$ transforming according to the
fundamental representation of $SU(N)\otimes U(1)$ and, 
to ensure anomaly cancellation, a 
multiplet $\{\phitil,\psitil,\Ftil\}$ transforming according to its conjugate.
The change $C^{\mu\nu}\go-C^{\mu\nu}$ for the conjugate representation is due 
to the fact that the anticommutation relations for the conjugate fundamental
representation 
differ by a sign from those for the fundamental representation. 
We denote the group matrices of $SU(N)\otimes U(1)$ by $R^A$ where
our convention is that $R^a$ are the $SU(N)$
generators and $R^0$ the $U(1)$ generator.  
The group matrices satisfy
\eqn\comrel{ [R^A,R^B]=if^{ABC}R^C,\quad
\{R^A,R^B\}=d^{ABC}R^C,}
where $d^{ABC}$ is totally symmetric. Of course then $f^{ABC}=0$ 
unless all indices are $SU(N)$. The matrices are normalised so that
$\Tr[R^AR^B]=\frak12\delta^{AB}$. In particular, $R^0=\sqrt{\frak{1}{2N}}1$.
We note that $d^{ab0}=\sqrt{\frak2N}\delta^{ab}$, $d^{000}=\sqrt{\frak2N}$. 
In Eq.~\lagranb, $C^{\mu\nu}$ is related to the non-anti-commutativity 
parameter $C^{\alpha\beta}$ by  
\eqn\Cmunu{
C^{\mu\nu}=C^{\alpha\beta}\epsilon_{\beta\gamma}
\sigma^{\mu\nu}_{\alpha}{}^{\gamma},} 
where 
\eqn\sigmunu{\eqalign{
\sigma^{\mu\nu}=&\frak14(\sigma^{\mu}\sigmabar^{\nu}-
\sigma^{\nu}\sigmabar^{\mu}),\cr
\sigmabar^{\mu\nu}=&\frak14(\sigmabar^{\mu}\sigma^{\nu}-
\sigmabar^{\nu}\sigma^{\mu}),\cr }} 
and 
\eqn\Csquar{
|C|^2=C^{\mu\nu}C_{\mu\nu}.} 
Our conventions are in accord with Ref.~\seib; in particular, 
\eqn\sigid{
\sigma^{\mu}\sigmabar^{\nu}=-\eta^{\mu\nu}+2\sigma^{\mu\nu}.}
Properties of $C$ which follow from
Eq.~\Cmunu\ are  
\eqna\cprop$$\eqalignno{
C^{\alpha\beta}&=\frak12\epsilon^{\alpha\gamma}
\left(\sigma^{\mu\nu}\right)_\gamma{}^{\beta}C_{\mu\nu},
& \cprop a\cr
C^{\mu\nu}\sigma_{\nu\alpha\betadot}&=C_{\alpha}{}^{\gamma}
\sigma^{\mu}{}_{\gamma\betadot},&\cprop b\cr
C^{\mu\nu}\sigmabar_{\nu}^{\alphadot\beta}&=-C^{\beta}{}_{\gamma}
\sigmabar^{\mu\alphadot\gamma}.&\cprop c\cr}$$ 

It is easy to show that Eq.~\lagranb\ is invariant under 
\eqn\newsusy{
\eqalign{
\delta A^A_{\mu}=&-i\lambdabar^A\sigmabar_{\mu}\epsilon\cr
\delta \lambda^A_{\alpha}=&i\epsilon_{\alpha}D^A+\left(\sigma^{\mu\nu}\epsilon
\right)_{\alpha}\left[F^A_{\mu\nu}
+\frak12iC_{\mu\nu}e^{ABC}d^{ABC}\lambdabar^B\lambdabar^C\right],\quad
\delta\lambdabar^A_{\alphadot}=0,\cr
\delta D^A=&-\epsilon\sigma^{\mu}D_{\mu}\lambdabar^A,\cr
\delta\phi=&\sqrt2\epsilon\psi,\quad\delta\phibar=0,\cr
\delta\psi^{\alpha}=&\sqrt2\epsilon^{\alpha} F,\quad 
\delta\psibar_{\alphadot}=-i\sqrt2(D_{\mu}\phibar)
(\epsilon\sigma^{\mu})_{\alphadot},\cr
\delta F=&0,\quad
\delta \Fbar=-i\sqrt2D_{\mu}\psibar\sigmabar^{\mu}\epsilon
-2i\phibar\epsilon\lambdahat+2C^{\mu\nu}D_{\mu}(\phibar\epsilon\sigma_{\nu}
\lambdahatbar).\cr}}
The terms involving $\gamma_{1-3}$ are 
separately invariant under $\Ncal=\frak12$ supersymmetry and must
be included to obtain a renormalisable lagrangian. In fact only the 
$\gamma_{1,2}$ terms were required in the case without a superpotential\jjwa;
to ensure renormalisability in the massive case we need to include
the $\gamma_3$ terms and also modify $\gamma_2$, with a corresponding change to
the bare gaugino $\lambda_B$ (see later).

We now consider the problem of adding superpotential terms to the 
lagrangian Eq.~\lagranb. This problem is most succinctly addressed by
returning to the superfield formalism whence the $\Ncal=\frak12$ action was
originally derived.
Denoting fundamental (antifundamental) chiral superfield representations as 
$\Phi$ ($\wt \Phi$)
it is simple to see that
$$
\int d^2 \theta \, \wt \Phi * \Phi \quad + \quad  
\int d^2 \bth\,\ov \Phi * \ov{\wt \Phi} 
$$
is gauge invariant, since under a gauge transformation we have 
$$
\Phi \to \Omega * \Phi, \quad \wt \Phi \to \wt \Phi *  \Omega^{-1}.
$$

In the \none\ case an interaction term is possible for the group $SU(3)$, 
i.e.
$$
\int d^2\theta\,\epsilon_{a b c} \Phi_1^a \Phi_2^b \Phi_3^c  
\quad + \quad\hbox{c.c.} 
$$
This construction does not, however, generalise to the \nhalf\ case, because 
of the non-anticommutative product. \nhalf\ invariant interaction terms 
{\it are\/} possible 
for chiral superfields in the adjoint representation; for a discussion of 
these see Ref.~\ref\jjwc{I.~Jack, D.R.T.~Jones and L.A.~Worthy,
``One-loop renormalisation of 
$\Ncal=\frak12$ supersymmetric gauge theory in the adjoint representation'',
in preparation}.

We may express the superfields in terms of component fields as follows:

\eqna\comps$$\eqalignno{
\Phi (y, \th) &= \phi (y) + \sqrt{2} \th \psi (y) + \th\th
F (y) & \comps a\cr
\wt \Phi (y, \th) &= \wt \phi  (y) + \sqrt{2} \th
\wt \psi (y) + \th\th \wt F  (y) & \comps b\cr
\ov \Phi (\ov y, \bth) &= \ov \phi (\ov y) + \sqrt{2} \bth \ov
\psi (\ov y) \cr
&\quad+ \bth\bth \Big( \ov F (\ov y) + i C^{\mu\nu} \partial_{\mu}
(\ov \phi  A_{\nu})(\ov y) -
\frak{1}{4} C^{\mu\nu} \ov \phi  A_{\mu} A_{\nu} (\ov y) \Big) & \comps c\cr
\ov {\wt \Phi}  (\ov y, \bth) &= \ov {\wt \phi}  (\ov y)
+ \sqrt{2} \bth \ov {\wt \psi}  (\ov y) \cr
&\quad+ \bth\bth \Big( \ov
{\wt F}  (\ov y) + i C^{\mu\nu} \partial_{\mu} (\ov {\wt \phi} 
A_{\nu})(\ov y) - \frak{1}{4} C^{\mu\nu} \ov {\wt \phi}  A_{\mu} A_{\nu} (\ov y)
\Big), & \comps d \cr}$$
where $\ov y^{\mu} = y^{\mu} - 2 i \th \sigma^{\mu} \bth$. Note the 
modification of the $\bth\bth$-term\araki.

We thus obtain

\eqna\expans$$\eqalignno{
m\int d^2 \theta \,\Phi * \wt \Phi  
&=  m\left[\phi \wt F + F \wt \phi  -\psi \wt\psi      \right] & \expans a \cr
m\int  d^2 \bth \,\ov \Phi * \ov{\wt \Phi} &=  m\left[\ov \phi \ov{\wt F} + 
\ov F \ov{\wt \phi} -\ov \psi \ov{\wt\psi}  
+ i C^{\mu\nu} \ov \phi {\hat F}_{\mu\nu}\ov{\wt\phi}\right] & \expans b \cr
}$$
In fact, the most general
mass term is in components
\eqn\lagmass{\eqalign{
S_{\rm{mass}}=&m\int d^4x\Bigl[(\phi\Ftil+F\phitil 
-\psi\psitil)
+\hbox{h.c.}+iC^{\mu\nu}\phibar \Fhat_{\mu\nu}\phitilbar\cr
&-\frak{1}{8}|C|^2d^{ABC}\phibar R^A\lambdahatbar^B\lambdahatbar^C
\phitilbar\Bigr].\cr}}
The coefficient
of the final term in Eq.~\lagmass\ is arbitrary since it is separately
$\Ncal=\frak12$ invariant; the reason for our particular choice will be
explained later (after Eq.~(A.6) in Appendix A).
This final term can also
be expressed in superfields but in a more  unwieldy form. 

%

We use the standard gauge-fixing term 
\eqn\gafix{
S_{\rm{gf}}={1\over{2\alpha}}\int d^4x (\pa.A)^2} 
with its associated
ghost terms.  The gauge propagators for $SU(N)$ and $U(1)$ are both given by  
\eqn\gprop{
\Delta_{\mu\nu}=-{1\over{p^2}}\left(\eta_{\mu\nu}
+(\alpha-1){p_{\mu}p_{\nu}\over{p^2}}\right)}
(omitting group factors) and the gaugino propagator is  
\eqn\fprop{
\Delta_{\alpha\alphadot}={p_\mu\sigma^{\mu}_{\alpha\alphadot}\over{p^2}},}
where the momentum enters at the end of the propagator with the undotted 
index.  
The one-loop graphs contributing
to the ``standard'' terms in the lagrangian (those without a
$C^{\mu\nu}$) are the same as in the ordinary $\Ncal=1$ case, so 
anomalous dimensions and gauge $\beta$-functions are as for
$\Ncal=1$. Since our gauge-fixing term in Eq.~\gafix\ does not preserve 
supersymmetry, the anomalous dimensions for $A_{\mu}$ and $\lambda$
are
different (and moreover gauge-parameter dependent), as are those for
$\phi$ and $\psi$. However, the 
gauge $\beta$-functions are of course gauge-independent. 
The one-loop one-particle-irreducible (1PI) 
graphs contributing to the new terms (those
containing $C$) in the absence of a superpotential were 
given in Ref.~\jjwa; the new diagrams in the presence of the 
mass terms are depicted in Figs.~1--3. The divergent contributions from these
diagrams are listed in Appendix A.

\newsec{Renormalisation of the $SU(N)\otimes U(1)$ action}
We found in Refs.~\jjw, \jjwa\ that non-linear renormalisations of 
$\lambda$ and $\Fbar$ were required to restore
gauge-invariance and ensure renormalisability at the one-loop level; and in 
a subsequent paper\ref\jjwb{I.~Jack, D.R.T.~Jones
and L.A.~Worthy, \prd72 (2005) 107701}\ we pointed out that non-linear 
renormalisations of $F$, $\Fbar$ are required even in ordinary $\Ncal=1$ 
supersymmetric gauge theory when working in the uneliminated formalism and
in the presence of a superpotential.
Note that in the $\Ncal=\frak12$ supersymmetric case, fields and their
conjugates may renormalise differently.
The renormalisations of the remaining fields and couplings are linear as 
usual and given by
\eqn\bare{\eqalign{ 
\lambdabar^a_B=Z_{\lambda}^{\frak12}\lambdabar^a,
\quad \lambdabar^0_B=Z_{\lambda^0}^{\frak12}\lambdabar^0,\quad&
A^{a}_{\mu B}=Z_A^{\frak12}A^{a}_{\mu},\quad A^{0}_{\mu B}
=Z_{A^0}^{\frak12}A^0_{\mu},\cr
D^a_B=Z_D^{\frak12}D^a,\quad& D^0_B=Z_{D^0}^{\frak12}D^0,\cr  
\phi_B=Z_{\phi}^{\frak12}\phi,
\quad \psi_B=Z_{\psi}^{\frak12}\psi,
\quad &\phibar_B=Z_{\phi}^{\frak12}\phibar,
\quad \psibar_B=Z_{\psi}^{\frak12}\psibar,\cr
\quad g_B=Z_gg,\quad g_{0B}=&Z_{g_0}g_0,
\quad m_B=Z_m m,\cr
\gamma_{1-3B}=Z_{1-3}, \quad 
C_B^{\mu\nu}=&Z_CC^{\mu\nu}, \quad |C|_B^2=Z_{|C|^2}|C|^2,
\cr}} 
with similar expressions for $\phitil_B$, $\psitil_B$ etc. (Note that
$F$ is unrenormalised in the absence of trilinear superpotential terms.)
In Eq.~\bare, $Z_{1-3}$ are divergent contributions, in other words
we have set the renormalised couplings $\gamma_{1-3}$ to zero for
simplicity. The other renormalisation constants start with 
tree-level values of 1. As we mentioned before,
the renormalisation constants for the fields
and for the gauge couplings $g$, $g_0$ are the same as in the ordinary $\Ncal=1$
supersymmetric theory and are therefore given up to one loop 
by\ref\timj{
D.~Gross and F.~Wilcek, \prd8 (1973) 3633\semi
D.R.T.~Jones, \npb87 (1975) 127}: 
\eqn\Zgg{\eqalign{
Z_{\lambda}=&1-g^2L(2\alpha N +2),\cr
Z_A=&1+g^2L[(3-\alpha)N-2]\cr
Z_D=&1-2g^2L,\cr
Z_g=&1+g^2L\left(1-3N\right),\cr
Z_{\phi}=&1+2(1-\alpha)L\Chat_2,\cr
Z_{\psi}=&1-2(1+\alpha)L\Chat_2,\cr
Z_m=&Z_{\Phi}^{-1},\cr
Z_{\Phi}=&1+4L\Chat_2\cr}}
where (using dimensional regularisation with $d=4-\epsilon$)
$L={1\over{16\pi^2\epsilon}}$ and
\eqn\ctildef{
\Chat_2=g^2R^aR^a+g_0^2R^0R^0=\frak12\left(Ng^2+\frak1N\Delta\right)}
with
\eqn\deltadefn{
\Delta=g_0^2-g^2.}
(For the gauge multiplet, 
we have given here the renormalisation constants corresponding to the
$SU(N)$ sector of the $U(N)$ theory; those for the $U(1)$ sector, namely
$Z_{\lambda^0}$, $Z_{A^0}$, $Z_{D^0}$
 and $Z_{g_0}$, are given by omitting the terms in
$N$ and replacing $g$ by $g_0$.) 
In Eq.~\Zgg, $Z_{\Phi}$ is the
renormalisation constant for the chiral superfield $\Phi$ so that the   
result for $m_B$ is the consequence of the non-renormalisation theorem.
For later convenience we write
(denoting for instance the $n$-loop
contribution to $Z_1$ by $Z_1^{(n)}$) 
\eqn\zforma{
Z^{(1)}_1=z_1L}
with similar expressions for $Z_{2,3}$.
The renormalisation of $\lambda^a$ is given by 
\eqn\lchange{\eqalign{
\lambda_B^a=&Z_{\lambda}^{\frak12}\lambda^a 
-\frak12NLg^3C^{\mu\nu}d^{abc}
\sigma_{\mu}\lambdabar^cA_{\nu}^b
-NLg^2g_0C^{\mu\nu}d^{ab0}
\sigma_{\mu}\lambdabar^0A_{\nu}^b\cr
&+i\sqrt2Lg\rho_1[\phibar R^a(C\psi)
+(C\psitil)R^a\bar{\tilde\phi}],\cr
\lambda_B^0=&iZ_{\lambda_0}^{\frak12}\lambda^0+
i\sqrt2Lg_0\rho_2[\phibar R^0(C\psi)+(C\psitil)R^0\bar{\tilde\phi}],\cr
}}
where $(C\psi)^{\alpha}=C^{\alpha}{}_\beta\psi^{\beta}$. Here $\rho_{4,5}$
are divergent parameters to be defined later.
The replacement
of $\lambda$ by $\lambda_B$ produces a change in the action given (to
first order) by
\eqn\Schlamb{\eqalign{
S_0(\lambda_B)-S_0(\lambda)=&L\int d^4x\Bigl\{
\rho_1\sqrt2gC^{\mu\nu}(\phibar\lambda^aR^a\sigmabar_{\nu}D_{\mu}\psi
+D_{\mu}\phibar\lambda^aR^a\sigmabar_{\nu}\psi)\cr
&+\rho_2\sqrt2g_0C^{\mu\nu}(\phibar\lambda^0R^0\sigmabar_{\nu}D_{\mu}\psi
+D_{\mu}\phibar\lambda^0R^0\sigmabar_{\nu}\psi)\cr
&+(\phi\go\phitil,\psi\go\psitil,R^A\go -(R^A)^*,C^{\mu\nu}\go
-C^{\mu\nu})+\ldots\Bigr\},\cr}}
where the ellipsis indicates the terms not involving $\rho_1$, $\rho_2$
(which were given previously in Ref.~\jjwa).

The final term in Eq.~\lagmass\ may be decomposed into four terms 
each of which are separately gauge and $\Ncal=\frak12$ invariant and
hence can (and do) renormalise separately. Consequently, in order to 
consider the renormalisation of the theory we need to replace Eq.~\lagmass\
by 
\eqn\lagmassnew{\eqalign{
S_{\rm{mass}}=&\int d^4x\Bigl\{m(\phi\Ftil+F\phitil
-\psi\psitil)
+\hbox{h.c.}+imC^{\mu\nu}\phibar \Fhat_{\mu\nu}\phitilbar\cr
&-\frak14|C|^2\phibar \Bigl(\frak12\mu_1g^2d^{abc}R^c\lambdabar^a\lambdabar^b
+\frak{1}{2N}\mu_2g^2\lambdabar^a\lambdabar^a\cr
&+2\mu_3gg_0R^aR^0\lambdabar^a\lambdabar^0
+\mu_4g_0^2R^0R^0\lambdabar^0\lambdabar^0\Bigr)
\phitilbar\Bigr\},\cr}}
where each of $\mu_{1-4}$ will renormalise separately.
However, for simplicity when we quote results for Feynman diagrams, we
use the values of the coefficents as implied by Eq.~\lagmass, i.e. $\mu_{1-4}
=m$; so that we are setting the renormalised values of $\mu_{1-4}$ to be $m$.

The redefinitions of $F$ and $\Fbar$ found in Ref.~\jjw\ need to be
modified in the presence of mass terms. This is easily done following the
arguments of Ref.~\jjwb; there are no one-loop diagrams giving divergent 
contributions to $m\phi F$ or $m\phibar\Fbar$ although there are counterterm
contributions from $m_B\phi_B F$, $m_B\phibar_B\Fbar$. However, note that due 
to the afore-mentioned change in sign for the $\phibar\lambdabar\lambdabar F$
term, the result for Fig.~8 in Ref.~\jjwb\ is modified to
\eqn\sumeight{\eqalign{
\Gamma_{8\rm{1PI}}^{(1)\rm{pole}}=
&L|C|^2\phibar\bigl\{g^2\left[\frak18(13-2\alpha)Ng^2-2\Chat_2\right]
\lambdabar^a\lambdabar^bd^{abc}R^c\cr
&+gg_0\left[\frak12(13-\alpha)Ng^2-8\Chat_2\right]
\lambdabar^0\lambdabar^aR^0R^a\cr
&-\left[2\Chat_2+\frak14\alpha Ng^2\right]
g^2d^{0bc}R^0\lambdabar^b\lambdabar^c
-4g_0^2\Chat_2\lambdabar^0\lambdabar^0R^0R^0\bigr\}F.
\cr}}      
We find
\eqn\fbarredef{\eqalign{
\Fbar_B=&\Fbar+(\alpha+3)mL\Chat_2\phitil+
L\Bigl\{\Bigl[\left(7Ng^2+2(1+\alpha) \Chat_2+2z_2\right)
g\pa_{\mu}A_{\nu}^a\cr
&-\left(\frak{15}{4}Ng^2+(1+\alpha)\Chat_2+z_2\right)
g^2f^{abc}A_{\mu}^bA_{\nu}^c\Bigr]iC^{\mu\nu}\phibar R^a\cr
&+2\left((1+\alpha)\Chat_2+z_3\right)
g_0\pa_{\mu}A_{\nu}^0iC^{\mu\nu}\phibar R^0\cr
&+\frak{1}{8}|C^2|\Bigl[\left(-19Ng^2+(17-\alpha)\Chat_2\right)
g^2d^{abc}\phibar R^c\lambdabar^a\lambdabar^b\cr
&+4\left(-16Ng^2+(17-\alpha)\Chat_2\right)
gg_0\phibar \lambdabar^0\lambdabar^aR^0R^a\cr
&+2(17-\alpha)\Chat_2
g_0^2\phibar \lambdabar^0\lambdabar^0R^0R^0
+\left(-6Ng^2+(17-\alpha)\Chat_2\right)g^2
d^{ab0}\phibar R^0\lambdabar^a\lambdabar^b\Bigr]\Bigr\},\cr
F_B=&F+(\alpha+3)mL\Chat_2\phitilbar.\cr}}
We now find that with
\eqn\zforms{\eqalign{
Z^{(1)}_C=&Z_{|C|^2}^{(1)}=0,\quad z_1=-3Ng^2, \quad
z_2=8(2\Chat_2-Ng^2), \cr
z_3=&4\left(\left[4-2\frak{\Delta}{g_0^2}\right]
\Chat_2-Ng^2\right),
\quad \rho_1=1+z_2,\quad  \rho_2=z_3, 
\cr
Z_{\mu_1}=&1+\left[\left(44+64\frak{g^2}{g_0^2}\right)\Chat_2
-\left(28+32\frak{g^2}{g_0^2}\right)Ng^2\right]L,\cr
Z_{\mu_2}=&1+\left[\left(44+128\frak{g^2}{g_0^2}\right)\Chat_2
-\left(28+32\frak{g^2}{g_0^2}\right)Ng^2\right]L,\cr
Z_{\mu_3}=&1+(44\Chat_2-30Ng^2)L,\cr
Z_{\mu_4}=&1+44\Chat_2L,\cr
\cr
}}
the one-loop effective action is finite. (In fact it is finite for
arbitrary choices of $z_2$, $z_3$; the particular values chosen are necessary 
to ensure renormalisability of the mass terms in Eq.~\lagmassnew.)

\newsec{The eliminated formalism}
It is instructive and also provides a useful check to perform the calculation 
in the eliminated formalism. In the eliminated case Eq.~\lagmassnew\ is replaced
by
\eqn\lagmassa{\eqalign{
\tilde S_{\rm{mass}}=&\int d^4x\Bigl\{m^2(\phibar\phi+\phitil\phitilbar)
-m(\psi\psitil+\psitilbar\psibar)\cr
&
-imC^{\mu\nu}\phibar \left[(1-2\gamma_2)gF^a_{\mu\nu}R^a
+(1-2\gamma_3)g_0F^0_{\mu\nu}R^0\right]\phitilbar\cr
&-\frak14|C|^2\phibar 
\Bigl(\frak12(\mu_1-2m)g^2d^{abc}R^c\lambdabar^a\lambdabar^b 
+\frak{1}{2N}(\mu_2-2m)g^2\lambdabar^a\lambdabar^a\cr
&+2(\mu_3-2m)gg_0R^aR^0\lambdabar^a\lambdabar^0
+(\mu_4-2m)g_0^2R^0R^0\lambdabar^0\lambdabar^0\Bigr)
\phitilbar\Bigr\}.\cr}}
while we simply strike out the terms involving $F$, $\Fbar$ in Eq.~\lagranb.
In Table~1, the contributions from Figs.~1(f-k) are now absent while those from
Figs.~1(l-r) change sign. Similarly, in Table~2, the contributions from
Figs.~2(e-p) are now absent while those from Figs.~2(q-dd) change sign.
In Table~3, the contributions from Figs.~3(f-n) are now absent while
those from Figs.~3(o-z) change sign. 
The results from Figs.~1, 2 and 3 now add to
\eqn\sumelim{\eqalign{
\Gamma_{1\rm{1PIelim}}^{(1)\rm{pole}}=&
iLgC^{\mu\nu}\pa_{\mu}A_{\nu}^A\phibar R^A\Bigl[\left\{-(68+4\alpha)
+32\frak{\Delta}{g_0^2}\delta^{A0}\right\}
\Chat_2\cr
&+\left\{(29-\alpha)c^A+16\delta^{A0}\right\}Ng^2\Bigr]
\phitilbar,\cr
\Gamma_{2\rm{1PIelim}}^{(1)\rm{pole}}=&
iLNg^2C^{\mu\nu}f^{abc}A_{\mu}^aA_{\nu}^b\phibar 
\left[2(17+\alpha)\Chat_2-(13-\alpha)Ng^2\right]R^c\phitilbar,\cr
\Gamma_{3\rm{1PIelim}}^{(1)\rm{pole}}=&\phibar\Bigl(
\left\{\left[\frak14(25+\alpha)+8\frak{g^2}{g_0^2}
\right]\Chat_2-\left[\frak14(11-\alpha)
+4\frak{g^2}{g_0^2}\right]Ng^2\right\}g^2d^{abc}R^c\lambdabar^a\lambdabar^b\cr
+&\left\{\left[\frak14(25+\alpha)+16\frak{g^2}{g_0^2}
\right]\Chat_2-\left[\frak14(11-\alpha)
+4\frak{g^2}{g_0^2}\right]Ng^2\right\}
\frak1Ng^2\lambdabar^a\lambdabar^a\cr
&+\left\{(25+\alpha)\Chat_2-\frak12(27-\alpha)g^2N\right\}gg_0R^aR^0
\lambdabar^a\lambdabar^0\cr
&+\frak12(25+\alpha)g_0^2\Chat_2R^0R^0\lambdabar^0\lambdabar^0\Bigr)
\phitilbar,\cr
}}
where $c^A=1-\delta^{A0}$.
The results in Eq.~\zforms\ are unchanged, which is a very good
check on the calculation.
 
\newsec{Conclusions}
We have constructed a set of mass terms for the $\Ncal=\frak12$ supersymmetric
theory with chiral matter in the fundamental representation, and we have shown 
that the $\Ncal=\frak12$ supersymmetry is preserved under renormalisation at
the one-loop level. However the renormalisability is assured by making a 
particular choice of the parameters $\gamma_2$, $\gamma_3$ 
(in Eq.~\lagranb) combined with a particular choice of renormalisations
for the gaugino 
$\lambda$, parametrised by $\rho_1$, $\rho_2$ (in Eq.~\lchange). These choices
were listed in Eq.~\zforms. This seems somewhat counterintuitive as these
renormalisations are all present in the massless theory and yet there
appeared to be nothing in the massless theory to enforce these choices.
It would be reassuring if some independent confirmation could be found for
these particular values. Presumably the necessity for the non-linear
renormalisations we are compelled to make lies in our use of a 
non-supersymmetric gauge (the obvious choice when working in components, of 
course). So the answer to this puzzle might lie in a close scrutiny of the
gauge-invariance Ward identities. Of course a calculation in superspace 
would also be illuminating. It is always tempting to investigate whether 
the behaviour at one loop persists to higher orders but the proliferation
of diagrams in this case would almost certainly be prohibitive.

\centerline{{\bf Acknowledgements}}\nobreak

LAW was supported by PPARC through a Graduate Studentship. 

\appendix{A}{Results for one-loop diagrams}
In this Appendix we list the divergent contributions from the 
one-loop diagrams. 

The divergent contributions
to the effective action from the graphs in Fig. 1 are of the form
\eqn\formone{
imLg^AC^{\mu\nu}\pa_{\mu}A_{\nu}^A\phibar R^AX_1^A\phitilbar}
where the contributions to $X_1^A$ from the
individual graphs are given in Table 1:
\bigskip
\vbox{
\begintable
Graph|$X_1^A$\cr
1a|$-4(2\Chat_2-Ng^2c^A)$\cr
1b|$-2Ng^2c^A$\cr
1c|$2(2\Chat_2-Ng^2c^A)$\cr
1d|$-4\Chat_2$\cr
1e|$-32\left(2-\frak{\Delta}{g_0^2}\delta^{A0}\right)\Chat_2
+(24-8\delta^{A0})Ng^2$\cr
1f|$-4\alpha(2\Chat_2-Ng^2c^A)$\cr
1g|$-4(2\Chat_2-Ng^2c^A)$\cr
1h|$4\Chat_2-Ng^2c^A$\cr
1i|$2(2\Chat_2-Ng^2c^A)$\cr
1j|$-2(1+2\alpha)Ng^2c^A$\cr
1k|$3Ng^2c^A$\cr
1l|$-(5+\alpha)Ng^2c^A$\cr
1m|$2\alpha(2\Chat_2-Ng^2c^A)$\cr
1n|$2(2\Chat_2-Ng^2c^A)$\cr
1o|$-(4\Chat_2-Ng^2c^A)$\cr
1p|$-2(2\Chat_2-Ng^2c^A)$\cr
1q|$2(1+2\alpha)Ng^2c^A$\cr
1r|$-3Ng^2c^A$
\endtable}
\centerline{{\it Table~1:\/} Contributions from Fig.~1}
\bigskip
These results add to
\eqn\sumone{\eqalign{
\Gamma_{1\rm{1PI}}^{(1)\rm{pole}}=&
imLgC^{\mu\nu}\pa_{\mu}A_{\nu}^A\phibar R^A\Bigl[\left\{-(76+4\alpha)
+32\frak{\Delta}{g_0^2}\delta^{A0}\right\}
\Chat_2\cr
&+\left\{(21+\alpha)c^A+16\delta^{A0}\right\}Ng^2\Bigr]
\phitilbar.\cr}}
(Note that the contributions from Figs.~1(h-k) cancel those from
Figs.~1(o-r) respectively.)

The divergent contributions
to the effective action from the graphs in Fig. 2 are of the form
\eqn\formseven{
imLg^2C^{\mu\nu}X_2f^{abc}A_{\mu}^aA_{\nu}^b\phibar R^c 
\phitilbar} 
where the contributions to $X_2$ from the
individual graphs are given in Table 2:
\bigskip 
\vbox{   
\begintable
Graph|$X_2$\cr
2a|$4(2\Chat_2-Ng^2)$\cr
2b|$2Ng^2$\cr
2c|$-2(2\Chat_2-Ng^2)$\cr
2d|$32\Chat_2-12Ng^2$\cr
2e|$2\alpha(2\Chat_2-Ng^2)$\cr
2f|$2(2\Chat_2-Ng^2)$\cr
2g|$-\frak12(3+\alpha)Ng^2$\cr
2h|$0$\cr
2i|$0$\cr
2j|$-2(2\Chat_2-Ng^2)$\cr
2k|$\frak32\alpha Ng^2$\cr
2l|$-\alpha Ng^2$\cr
2m|$\frak32\alpha Ng^2$\cr
2n|$\frak12(2+\alpha)Ng^2$\cr
2o|$0$\cr
2p|$0$\cr
2q|$-\frak32\alpha Ng^2$\cr
2r|$\frak32(1+\alpha)Ng^2$\cr
2s|$-\alpha(2\Chat_2-Ng^2)$\cr
2t|$-(2\Chat_2-Ng^2)$\cr
2u|$\frak12(3+\alpha)Ng^2$\cr
2v|$0$\cr
2w|$0$\cr
2x|$2(2\Chat_2-Ng^2)$\cr
2y|$-\frak32\alpha Ng^2$
\endtable}
\centerline{{\it Table~2:\/} Contributions from Fig.~2}
\vbox{
\begintable
Graph|$X_2$\cr
2z|$\alpha Ng^2$\cr
2aa|$-\frak32\alpha Ng^2$\cr
2bb|$-\frak12(2+\alpha)Ng^2$\cr
2cc|$0$\cr
2dd|$0$
\endtable}
\centerline{{\it Table~2:\/} Contributions from Fig.~2 (continued)}
\bigskip 
These results add to
\eqn\sumtwo{
\Gamma_{2\rm{1PI}}^{(1)\rm{pole}}=
imLNg^2C^{\mu\nu}f^{abc}A_{\mu}^aA_{\nu}^b\phibar 
\left[2(19+\alpha)\Chat_2-(\frak{23}{2}+\alpha)Ng^2\right]R^c\phitilbar}
(Note that the contributions from Figs.~2(g-p) cancel those from
Figs.~2(u-dd) respectively.)

The divergent contributions
to the effective action from the graphs in Fig. 3 are of the form
\eqn\formthree{
mL|C|^2g^Ag^B\lambdabar^A\lambdabar^B\phibar X_3^{AB}
\phitilbar}
where the contributions to $X_3^{AB}$ from the
individual graphs are given in Table 3.
The results from Table 3 add to 
\eqn\sumthree{\eqalign{
\Gamma_{3\rm{1PI}}^{(1)\rm{pole}}=&mL|C|^2\phibar\Bigl(
\left\{\left[\frak14(2-\alpha)-4\frak{g^2}{g_0^2}\right]Ng^2
+\left[\frak14(9+\alpha)+\frak{8g^2}{g_0^2}
\right]\Chat_2\right\}g^2d^{abc}R^c\lambdabar^a\lambdabar^b\cr
+&\left\{-\left[\frak14(11+\alpha)+4\frak{g^2}{g_0^2}\right]Ng^2
+\left[\frak14(9+\alpha)+\frak{16g^2}{g_0^2}
\right]\Chat_2\right\}
\frak1Ng^2\lambdabar^a\lambdabar^a\cr
&+\left\{(9+\alpha)\Chat_2-\frak12(1+\alpha)Ng^2\right\}gg_0R^aR^0
\lambdabar^a\lambdabar^0\cr
&+\frak12(9+\alpha)g_0^2\Chat_2R^0R^0\lambdabar^0\lambdabar^0\Bigr)
\phitilbar.\cr}}
(Note that the contributions from Figs.~3(h--m) cancel those from 
Figs.~3(u--z), in analogy to the situation with Figs.~1 and 2; this is a 
consequence of our choice of coefficient for the last term in Eq.~\lagmass.)
\bigskip
\vbox{
\begintable
Graph|$X_3^{ab}$|$X_3^{a0}$|$X_3^{00}$\cr
3a|$\alpha\left(\frak{1}{2N}\Delta R^aR^b+\frak14g^2\delta^{ab}\right)$|
$\alpha\frak{1}{2N}\Delta R^aR^0$|$\alpha \Chat_2R^0R^0$\cr
3b|$\left(\frak{1}{2N}\Delta R^aR^b+\frak14g^2\delta^{ab}\right)$|
$\frak{1}{2N}\Delta R^aR^0$|$\Chat_2R^0R^0$\cr
3c|$(3+\alpha)\left(\frak{1}{2N}\Delta R^aR^b+\frak14g^2\delta^{ab}\right)$|
$(3+\alpha)\frak{1}{2N}\Delta R^aR^0$|$(3+\alpha) \Chat_2R^0R^0$\cr
3d|$-2\alpha\left(\frak{1}{2N}\Delta R^aR^b+\frak14g^2\delta^{ab}\right)$|
$-2\alpha\frak{1}{2N}\Delta R^aR^0$|$-2\alpha \Chat_2R^0R^0$\cr
3e|$\left(Ng^2+4\frak{g^2\Delta}{g_0^2N}\right)d^{abc}R^c+\frak{2g^2}{g_0^2}
\left(2g^2-g_0^2+4\frak{\Delta}{N^2}\right)\delta^{ab}$|$0$|$0$\cr
3f|$\alpha\left(\frak{1}{2N}\Delta R^aR^b+\frak14g^2\delta^{ab}\right)$|
$\alpha\frak{1}{2N}\Delta R^aR^0$|$\alpha \Chat_2R^0R^0$\cr
3g|$\left(\frak{1}{2N}\Delta R^aR^b+\frak14g^2\delta^{ab}\right)$|
$\frak{1}{2N}\Delta R^aR^0$|$\Chat_2R^0R^0$\cr 
3h|$-2\left(\frak{1}{N}\Delta R^aR^b+\frak14g^2\delta^{ab}\right)$|
$-\frak12\left(2\Chat_2+\frak3N\Delta\right)R^aR^0$|$-4\Chat_2R^0R^0$\cr
3i|$\frak14\alpha g^2 Nd^{abc}R^c$|$\frak12\alpha NR^aR^0$|$0$\cr
3j|$-\alpha g^2Nd^{abc}R^c$|$-2\alpha NR^aR^0$|$0$\cr      
3k|$8\left(\frak{1}{N}\Delta R^aR^b+\frak14g^2\delta^{ab}\right)$|
$2\left(2\Chat_2+\frak3N\Delta\right)R^aR^0$|$16\Chat_2R_0R_0$\cr
3l|$\alpha g^2 Nd^{abc}R^c$|$2\alpha NR^aR^0$|$0$\cr
3m|$-4\left(\frak{1}{2N}\Delta R^aR^b+\frak14g^2\delta^{ab}\right)$|     
$-\frak{2}{N}\Delta R^aR^0$|$-4\Chat_2R^0R^0$\cr             
3n|$0$|$0$|$0$\cr
3o|$0$|$0$|$0$\cr   
3p|$-\alpha g^2 Nd^{abc}R^c$|$-2\alpha NR^aR^0$|$0$\cr
3q|$(1+\alpha) g^2 Nd^{abc}R^c$|$2(1+\alpha) NR^aR^0$|$0$\cr
3r|$-\frak18(3+\alpha)g^2(Nd^{abc}R^c+2g^2\delta^{ab})$|$0$|$0$\cr
3s|$-\frak12\alpha\left(\frak{1}{2N}\Delta R^aR^b+\frak14g^2\delta^{ab}\right)$|
$-\frak12\alpha\frak{1}{2N}\Delta R^aR^0$|$-\frak12\alpha \Chat_2R^0R^0$\cr
3t|$-\frak12\left(\frak{1}{2N}\Delta R^aR^b+\frak14g^2\delta^{ab}\right)$|
$-\frak12\frak{1}{2N}\Delta R^aR^0$|$-\frak12 \Chat_2R^0R^0$\cr
3u|$2\left(\frak{1}{N}\Delta R^aR^b+\frak14g^2\delta^{ab}\right)$|
$\frak12\left(2\Chat_2+\frak3N\Delta\right)R^aR^0$|$4\Chat_2R^0R^0$\cr
3v|$-\frak14\alpha g^2 Nd^{abc}R^c$|$-\frak12\alpha NR^aR^0$|$0$\cr
3w|$\alpha g^2Nd^{abc}R^c$|$2\alpha NR^aR^0$|$0$\cr
3x|$-8\left(\frak{1}{N}\Delta R^aR^b+\frak14g^2\delta^{ab}\right)$|
$-2\left(2\Chat_2+\frak3N\Delta\right)R^aR^0$|$-16\Chat_2R_0R_0$
\endtable}
\centerline{{\it Table~3:\/} Contributions from Fig.~3}
\vfill
\eject
\vbox{
\begintable
Graph|$X_3^{ab}$|$X_3^{a0}$|$X_3^{00}$\cr
3y|$-\alpha g^2 Nd^{abc}R^c$|$-2\alpha NR^aR^0$|$0$\cr
3z|$4\left(\frak{1}{2N}\Delta R^aR^b+\frak14g^2\delta^{ab}\right)$|
$\frak{2}{N}\Delta R^aR^0$|$4\Chat_2R^0R^0$\cr
3aa|$0$|$0$|$0$\cr
3bb|$0$|$0$|$0$
\endtable}
\centerline{{\it Table~3:\/} Contributions from Fig.~3 (continued)}
\bigskip
\vfill
\eject  
\epsfysize= 7in
\centerline{\epsfbox{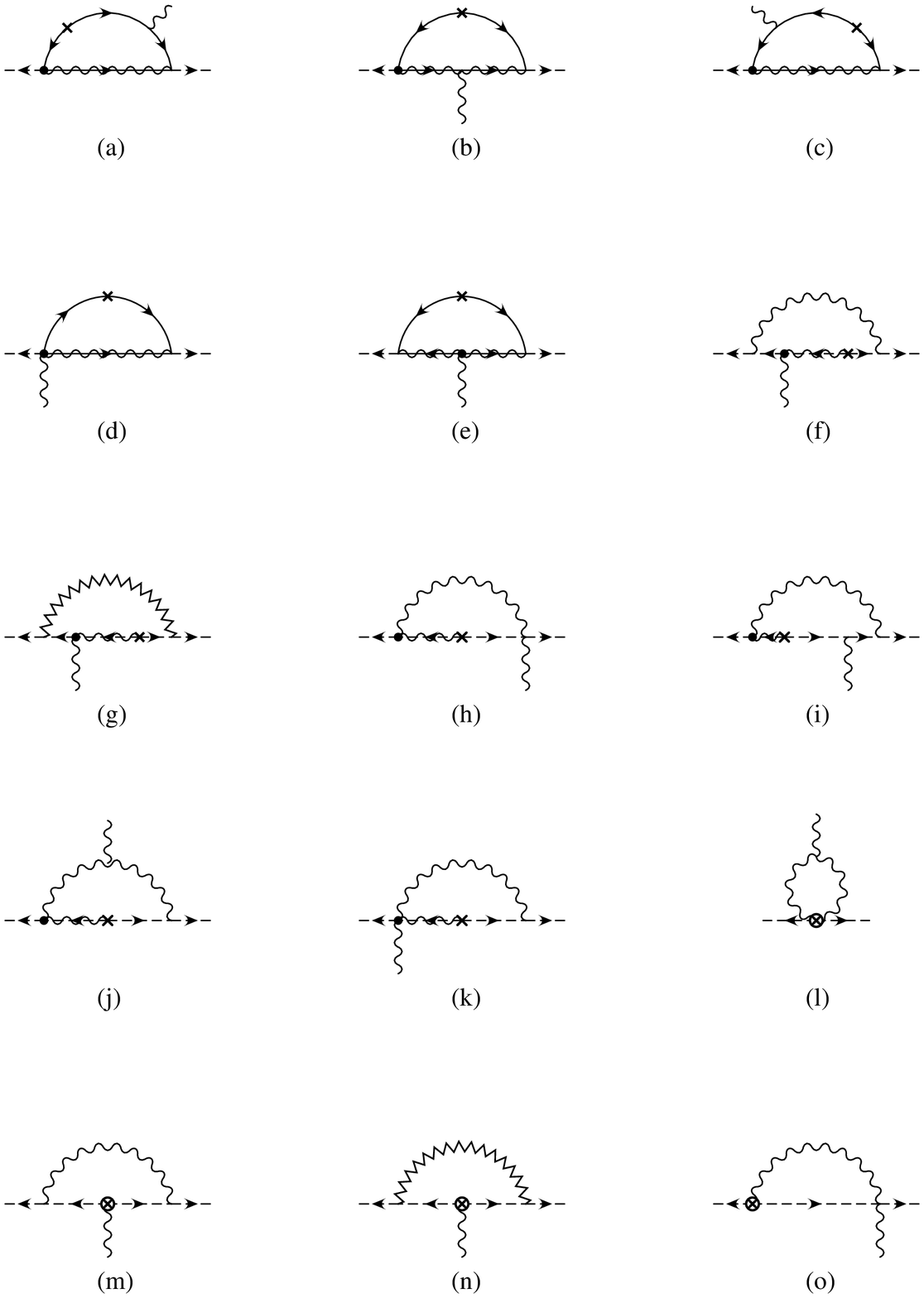}}
\inparg 
{\it \noindent Fig. 1: Diagrams with two scalar,
one gauge lines; a dot denotes a $C$, a cross a mass
and a crossed circle a vertex with both a mass and a $C$.}
\medskip
\outparg

\bigskip
\epsfysize= 1.2in
\centerline{\epsfbox{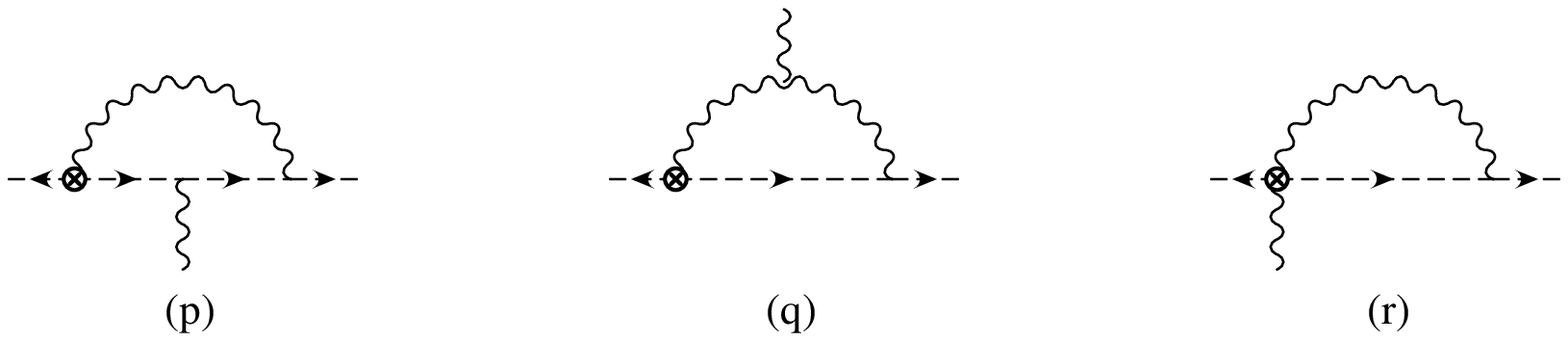}}
\inparg
{\it \noindent Fig. 1(continued)}
\medskip
\outparg
\vfill
\eject  
\epsfysize= 7in
\centerline{\epsfbox{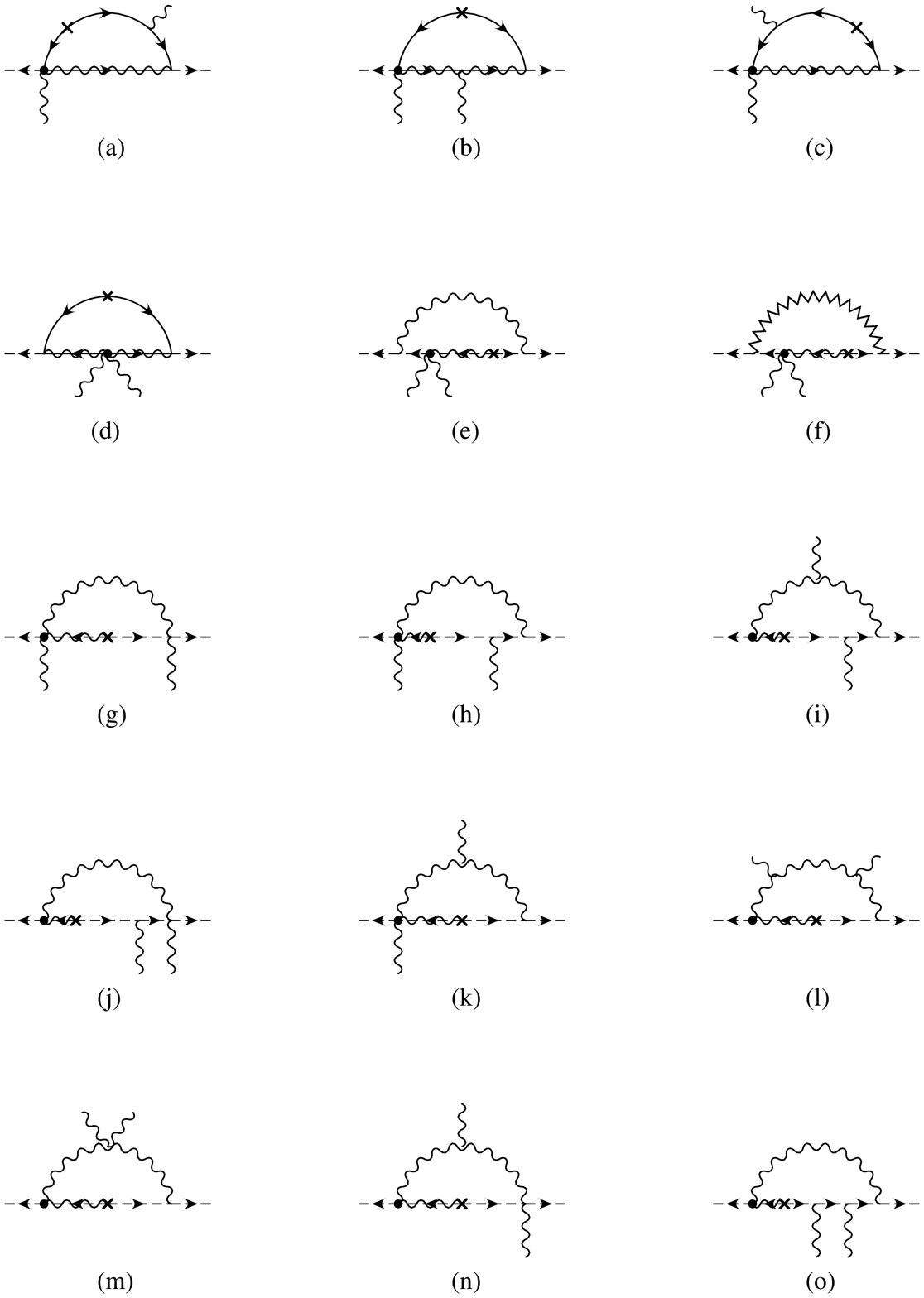}}
\inparg 
{\it \noindent Fig. 2: Diagrams with two scalar, two
gauge lines; a dot denotes a $C$, a cross a mass
and a crossed circle a vertex with both a mass and a $C$.}
\medskip
\outparg
\vfill  
\eject  
\epsfysize= 7in
\centerline{\epsfbox{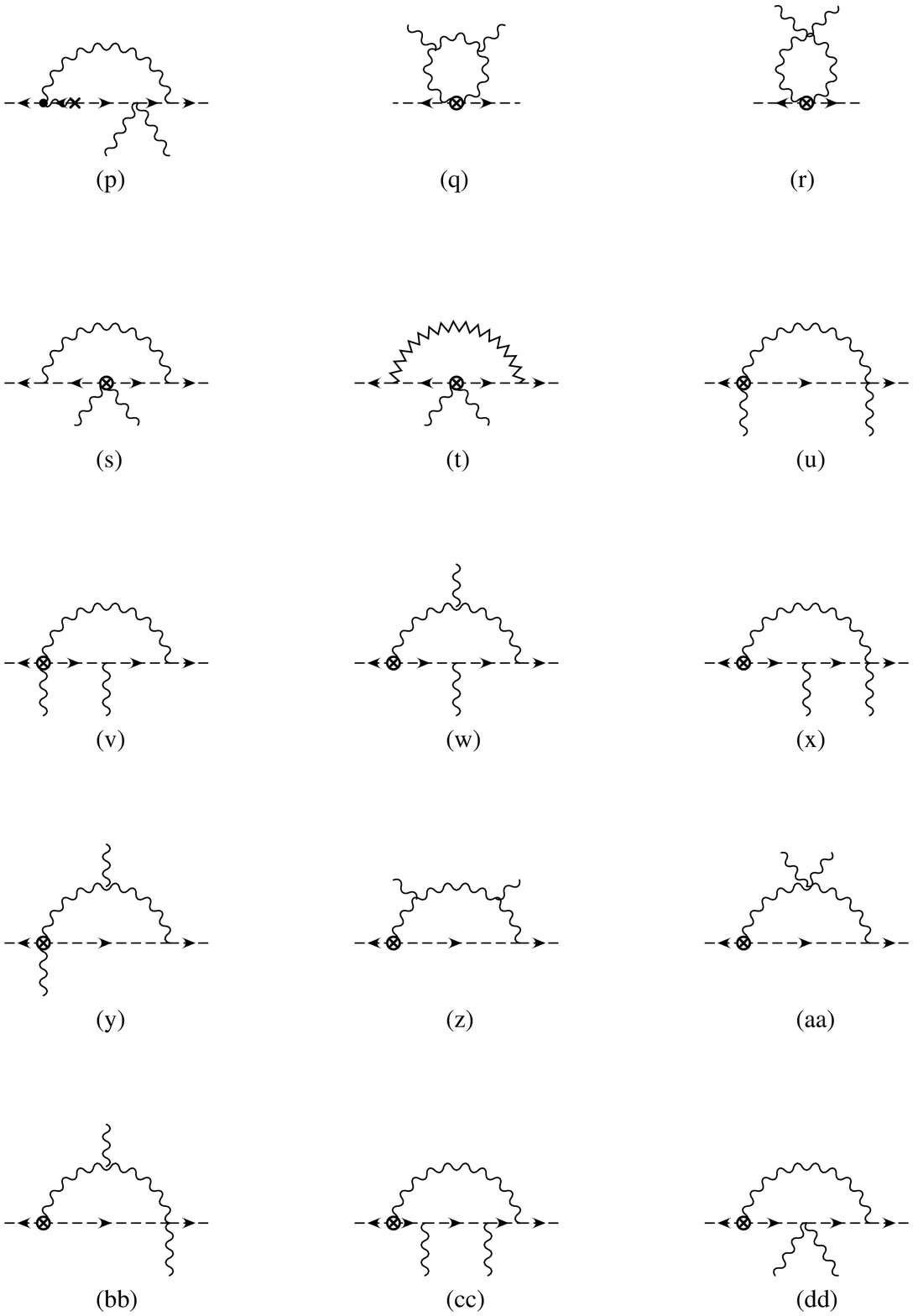}}
\inparg
{\it \noindent Fig. 2 (continued)}
\medskip
\outparg

\vfill  
\eject  
\epsfysize= 7in
\centerline{\epsfbox{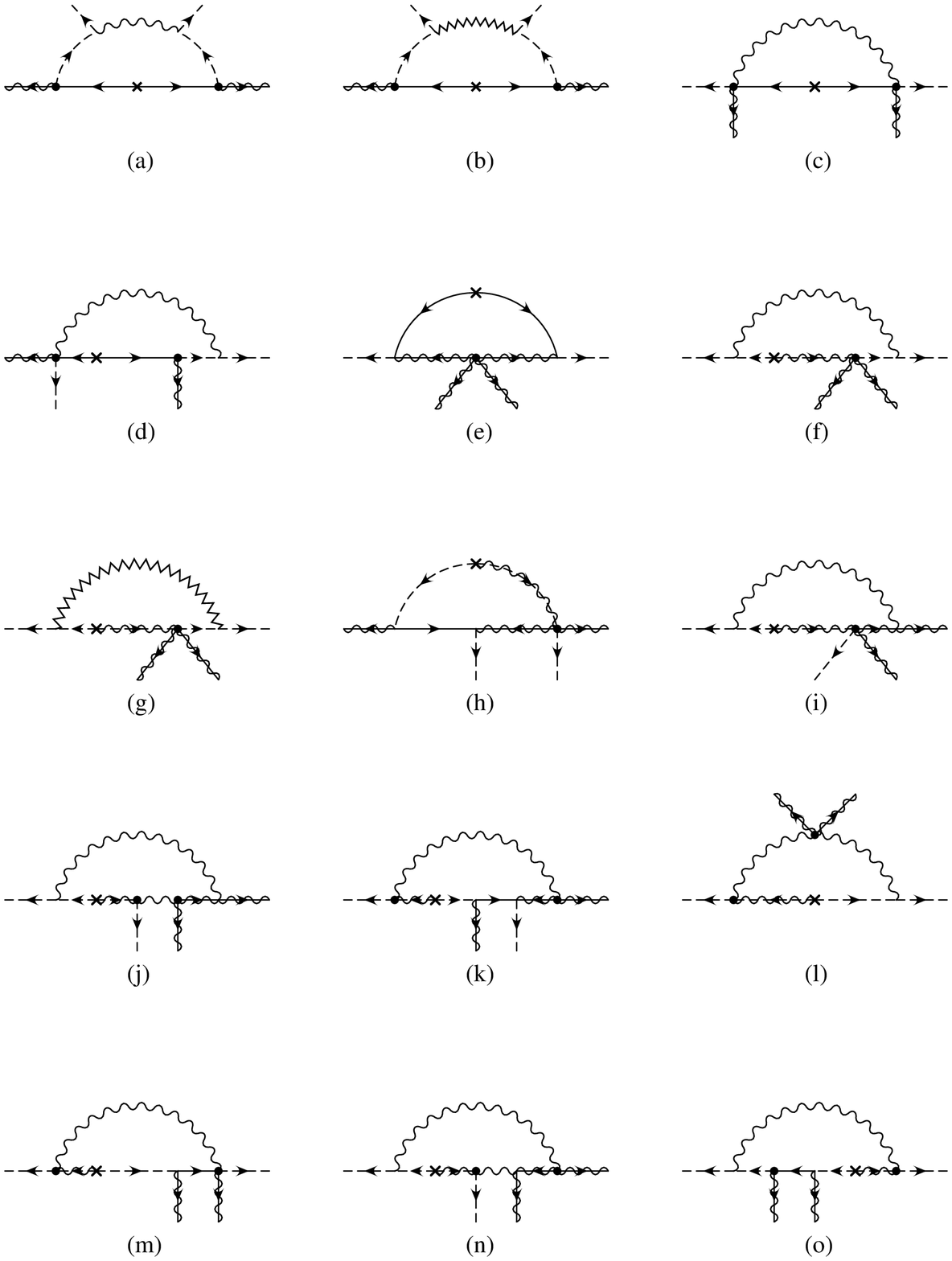}}
\inparg
{\it \noindent Fig. 3: Diagrams with two scalar, two gaugino lines;
a dot denotes a $C$, a cross a mass
and a crossed circle a vertex with both a mass and a $C$.}
\medskip
\outparg
\vfill  
\eject
\epsfysize= 7in
\centerline{\epsfbox{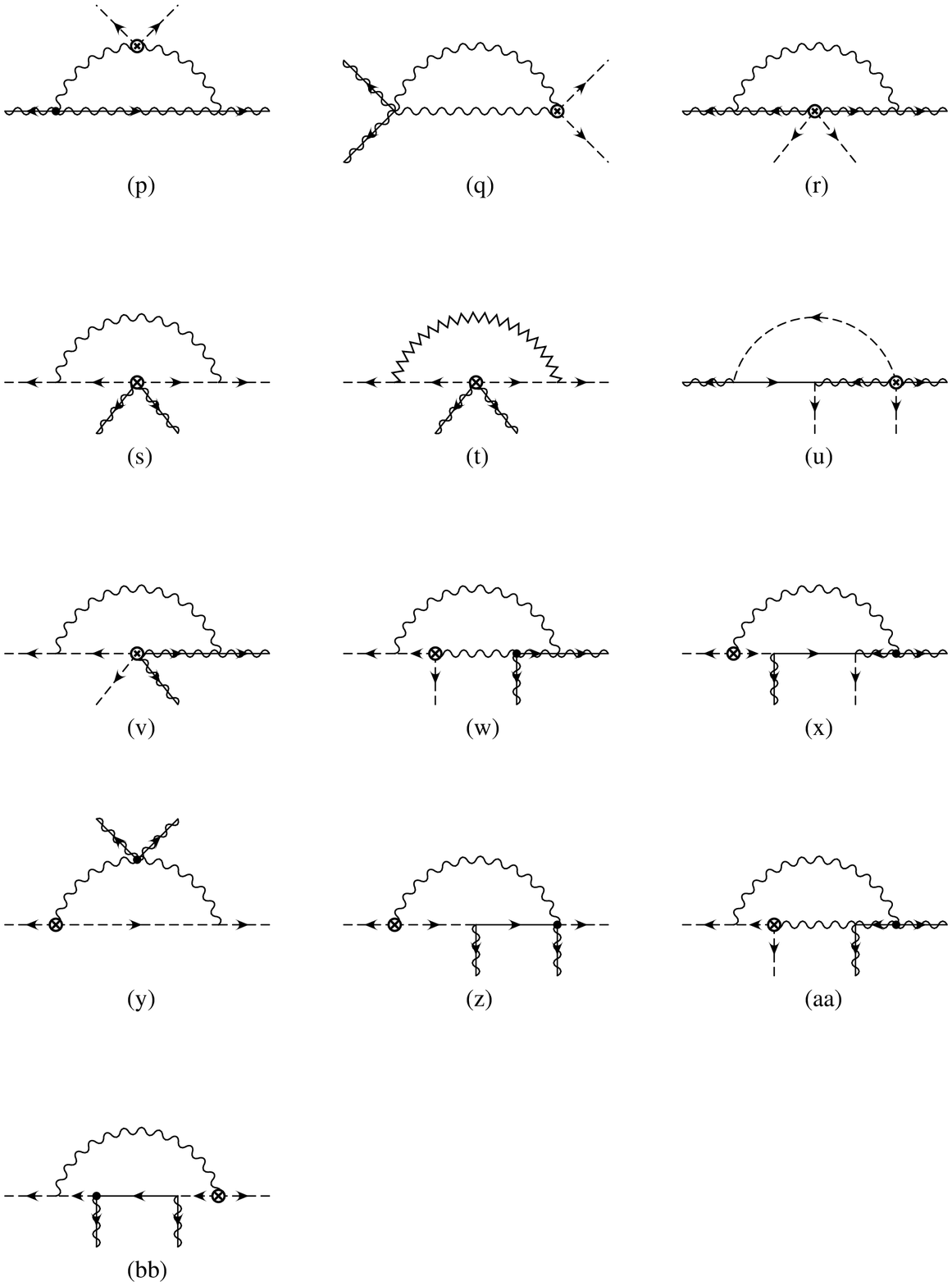}}
\inparg
{\it \noindent Fig. 3 (continued)}
\medskip
\outparg
\vfill
\eject  
\listrefs
\bye